\documentclass[twocolumn,aps,superscriptaddress]{revtex4-1}

\usepackage{graphicx}% Include figure files
\usepackage{dcolumn}% Align table columns on decimal point
\usepackage{bm}% bold math
\usepackage[cmex10]{amsmath}
\usepackage[utf8]{inputenc}
\usepackage[T1]{fontenc}
\usepackage{natbib,hyperref}
\usepackage{siunitx}

\begin{document}

\preprint{APS/123-QED}

\title{Buffer influence on magnetic dead layer, critical current and thermal stability in magnetic tunnel junctions with perpendicular magnetic anisotropy}

\author{Marek Frankowski}
 \email{mfrankow@agh.edu.pl}
 \affiliation{AGH University of Science and Technology, Department of Electronics, Al. Mickiewicza 30, 30-059 Krak\'{o}w, Poland}
 \author{Antoni \.{Z}ywczak}
 \email{zywczak@agh.edu.pl}
\affiliation{AGH University of Science and Technology, Academic Center of Materials and Nanotechnology, Al. Mickiewicza 30, 30-059 Krak\'{o}w, Poland}
\author{Maciej Czapkiewicz}
\author{S\l{}awomir Ziętek}
\author{Jaros\l{}aw Kanak}
\author{Monika Banasik}
\author{Wiesław Powroźnik}
\author{Witold Skowro\'{n}ski}
\author{Jakub Chęci\'{n}ski}
\affiliation{AGH University of Science and Technology, Department of Electronics, Al. Mickiewicza 30, 30-059 Krak\'{o}w, Poland}

\author{Jerzy Wrona}
\affiliation{AGH University of Science and Technology, Department of Electronics, Al. Mickiewicza 30, 30-059 Krak\'{o}w, Poland}
\affiliation{Singulus Technologies, Kahl am Main, 63796, Germany}

\author{Hubert Głowiński}
\author{Janusz Dubowik}
\affiliation{Institute of Molecular Physics, Polish Academy of Sciences, ul. Mariana Smoluchowskiego 17,
60-179 Poznań, Poland}

\author{Jean-Philippe Ansermet}
\affiliation{Ecole Polytechnique Fédérale de Lausanne, IPMC-Station 3, CH-1015 Lausanne-EPFL, Schwitzerland}

\author{Tomasz Stobiecki}
\affiliation{AGH University of Science and Technology, Department of Electronics, Al. Mickiewicza 30, 30-059 Krak\'{o}w, Poland}

%\author{G\"{u}nter Reiss}
%\affiliation{Thin Films and Physics of Nanostructures, Bielefeld University, 33615 Bielefeld, Germany}

%\author{Sebastiaan van Dijken}
%\affiliation{NanoSpin, Department of Applied Physics, Aalto University School of Science, P.O.Box 15100, FI-00076 Aalto, Finland}

\date{\today}% It is always \today, today,

\begin{abstract}
We present a thorough
research on Ta/Ru-based buffers and their influence on features crucial from the point of view of applications of MTJs, such as critical switching current and thermal stability.
We investigate devices consisting of buffer/FeCoB/MgO/FeCoB/Ta/Ru multilayers for three different buffers: Ta 5 / Ru 10 / Ta 3, Ta 5 /
Ru 10 / Ta 10 and Ta 5 / Ru 20 / Ta 5 (all thicknesses in nm). In addition, we study systems with a single FeCoB layer deposited above as well as below
the MgO barrier. The crystallographic texture and the roughness of the buffers are determined by means of XRD and atomic force
microscopy measurements. Furthermore, we examine the magnetic domain pattern, the magnetic dead layer thickness
and the perpendicular magnetic anisotropy fields for each sample. Finally, we investigate the effect of the current induced magnetization switching for nanopillar junctions with lateral dimensions ranging from 1 $\mu$m down to 140 nm. Buffer Ta 5 / Ru 10 / Ta 3, which has the thickest dead layer, exhibits a large increase in the thermal stability factor while featuring a slightly lower critical current density value when compared to the buffer with the thinnest dead layer Ta 5 / Ru 20 / Ta 5.

 %We compare results
%Transport
%properties have been investigated in nanopillars with lateral dimensions ranging from 1 $\mu$m down to 140 nm in
%diameter. We have determined magnetic dead layer thickness of Fe$_{60}$Co$_{20}$B$_{20}$ for each buffer.
%Sample with the third buffer exhibited the thickest dead layer and for this sample an optimal PMA has been
%reached, as the effective thickness of Fe$_{60}$Co$_{20}$B$_{20}$ layer was the smallest.
%Current induced magnetization switching (CIMS) experiment has been conducted showing the lowest intrinsic critical
%current density for sample with the first listed buffer.
\end{abstract}

%\pacs{75.47.-m, 72.25.-b}% PACS, the Physics and Astronomy
\maketitle	

\section{Introduction}
Magnetic Tunnel Junctions (MTJs) with Perpendicular Magnetic Anisotropy (PMA) have brought significant attention in view of numerous applications such as magnetic field sensors \cite{wisniowski2008effect,wei2009magnetic,wisniowski2013magnetic,wisniowski2014reduction}, microwave generators and detectors \cite{kiselev2003microwave,boulle2007shaped,deac2008bias,skowronski2014spin} and high-density non-volatile magnetic random access memory cells \cite{nishimura2002magnetic,ikeda2010perpendicular,meng2006spin,sato2011junction,sbiaa2011materials}. The latter is particularly interesting due to the advantageous features exhibited by MTJs which include low critical switching current density, good thermal stability, low power consumption and the ability to scale down the junction size \cite{nishimura2002magnetic,ikeda2010perpendicular,meng2006spin,sato2011junction}. In general, as the PMA can be affected by the MTJ layer structure \cite{tao2014perpendicular,barsukov2014field,cheng2011effect,chang2013effect,sokalski2012optimization}, these properties can also be modified significantly, creating an opportunity for further improvement of the magnetic memory technology based on MTJs. Recently, a lot of attention has been paid to layer thickness and buffer material problems in FeCoB/MgO systems \cite{barsukov2014field,cheng2011effect,chang2013effect,sokalski2012optimization,
sokalski2013increased}, which are widely used to achieve large Tunneling Magnetoresistance (TMR) values \cite{barsukov2014field,ikeda2008tunnel,ikeda2010perpendicular}. Different buffer layer textures may influence the roughness and thus the electrical and magnetic properties of the samples \cite{sort2005exchange,wisniowski2006influence,kanak2008influence}, affecting the parameters which are crucial in the context of magnetic memory. We discuss Ta/Ru buffers, which are used by the nanoelectronics industry \cite{gottwald2013ultra}. Moreover, buffers with Ta layers are particularly interesting, since they are also commonly used in Spin Hall Effect experiments \cite{liu2012spin,morota2011indication,hahn2013comparative}. 

In this work, we have used FeCoB/MgO MTJs with three different sets of Ta/Ru/Ta buffer layers in order to investigate the magnetic dead layer thickness, the critical current and the thermal stability. In Section \ref{sec:experimental}, we describe in details the preparation and the layer structure of the junctions as well as experimental methods used. As presented in Section \ref{sec:results}, we have performed wafer-level measurements to characterize MTJs structural and magnetic properties for different buffer types. By means of a Current Induced Magnetization Switching (CIMS) experiment conducted on patterned samples, the transport properties have also been investigated. The experimental results are discussed and the physical explanation for the observed differences between MTJ parameters measured with various buffer types is proposed. Finally, in Section \ref{sec:summary}, we present a summary and conclusions.

\section{Experimental\label{sec:experimental}}
The MTJ stack
has been deposited on a thermally oxidized silicon wafer (SiO$_2$ thickness 100 nm) using a Singulus Timaris cluster tool system with the multilayer structure as follows (all thicknesses in nm): buffer / Fe$_{60}$Co$_{20}$B$_{20}$ 1.0 / MgO wedge / Fe$_{60}$Co$_{20}$B$_{20}$ 1.5 / Ta 5 / Ru 5, for three different buffers: (a) Ta 5 / Ru 10 / Ta 3, (b) Ta 5 / Ru 10 / Ta 10 and (c) Ta 5 / Ru 20 / Ta 5.

We have also prepared two single ferromagnetic layer systems: one with a ferromagnetic layer deposited below the MgO layer (buffer / Fe$_{60}$Co$_{20}$B$_{20}$ wedge / MgO 1.28 / Ta 5 / Ru 5), which will be further referred to as \textit{bottom} and another one with a ferromagnetic layer deposited above the MgO layer (buffer / MgO 1.28 / Fe$_{60}$Co$_{20}$B$_{20}$ wedge / Ta 5 / Ru 5), which will be further referred to as \textit{top}. The measurements have been performed for samples before and after the thermal treatment of 330$^{\circ}$C for one hour with the perpendicular magnetic field bias of 0.42 T.

Crystallographic properties of the prepared samples have been investigated using XRD $\theta$ - $2\theta$ and rocking curve measurements. The surface topography and the grain size dependence on the buffer have been examined with an Atomic Force Microscope (AFM). Single layer systems have been investigated by means of a Vibrating Sample Magnetometer (VSM) and a polar Magnetooptic Kerr Effect (p-MOKE) microscope. The magnetization, the anisotropy fields and the magnetic dead layer thickness have been found using measurements for different FeCoB layer thicknesses.

%Tunnel magnetoresistance of MTJ with the selected MgO thickness of 1.05 nm
%for each buffer has been equal to (a) 102\%, (b) 94\% and (c) 88\%, taken from current-in-plane tunnelling in annealed samples.
The MTJs have been patterned into circular and elliptical shaped pillars with lateral dimensions ranging from $\SI{1 }{\micro\meter}$ down to 140 nm using an electron beam lithography, an ion-beam etching and a lift-off process. The CIMS experiment has been conducted with different current pulse time widths and the intrinsic critical currents and thermal stability factors have been calculated for the MTJs with each buffer \cite{kubota2005evaluation}. To determine the damping coefficients, we have used Ferromagnetic Resonance detected with a Vector Network Analyzer (VNA-FMR).

\section{Results and discussion\label{sec:results}}
\subsection{Microstructure: texture and roughness}
The lowermost Ta buffer layer deposited directly on SiO$_2$ was amorphous, which is in agreement with our previous investigations \cite{kanak2013x}, whereas the two remaining layers Ru and Ta were both highly textured.  Figure 1 shows XRD $\theta$ - $2\theta$ diffraction patterns in a narrow $2\theta$ range for the samples with (a), (b) and (c) buffers. Different intensities of Ru and Ta peaks result from different thicknesses of the layers. In wide angular 2$\theta$ range XRD measurements (not shown) only Ru (002) and Ta (110) peaks and their second order are visible, suggesting that Ru and Ta buffer layers are highly oriented. The Ru layer in the buffer has grown polycrystalline in a columnar structure, which is clearly visible in TEM images shown for similar buffer structures in Ref. \cite{kanak2013x,karthik2012transmission}. As has been shown in Ref.\cite{kanak2013x}, the XRD profiles and Monte Carlo simulations confirm the columnar growth of Ta/Ru/Ta buffers. Ru in these buffers has grown in hcp (002)-oriented texture whereas Ta has grown in bcc (110)-oriented texture. 

            \begin{figure}[!htbp]
            \centering
            \includegraphics[width=0.5\textwidth]{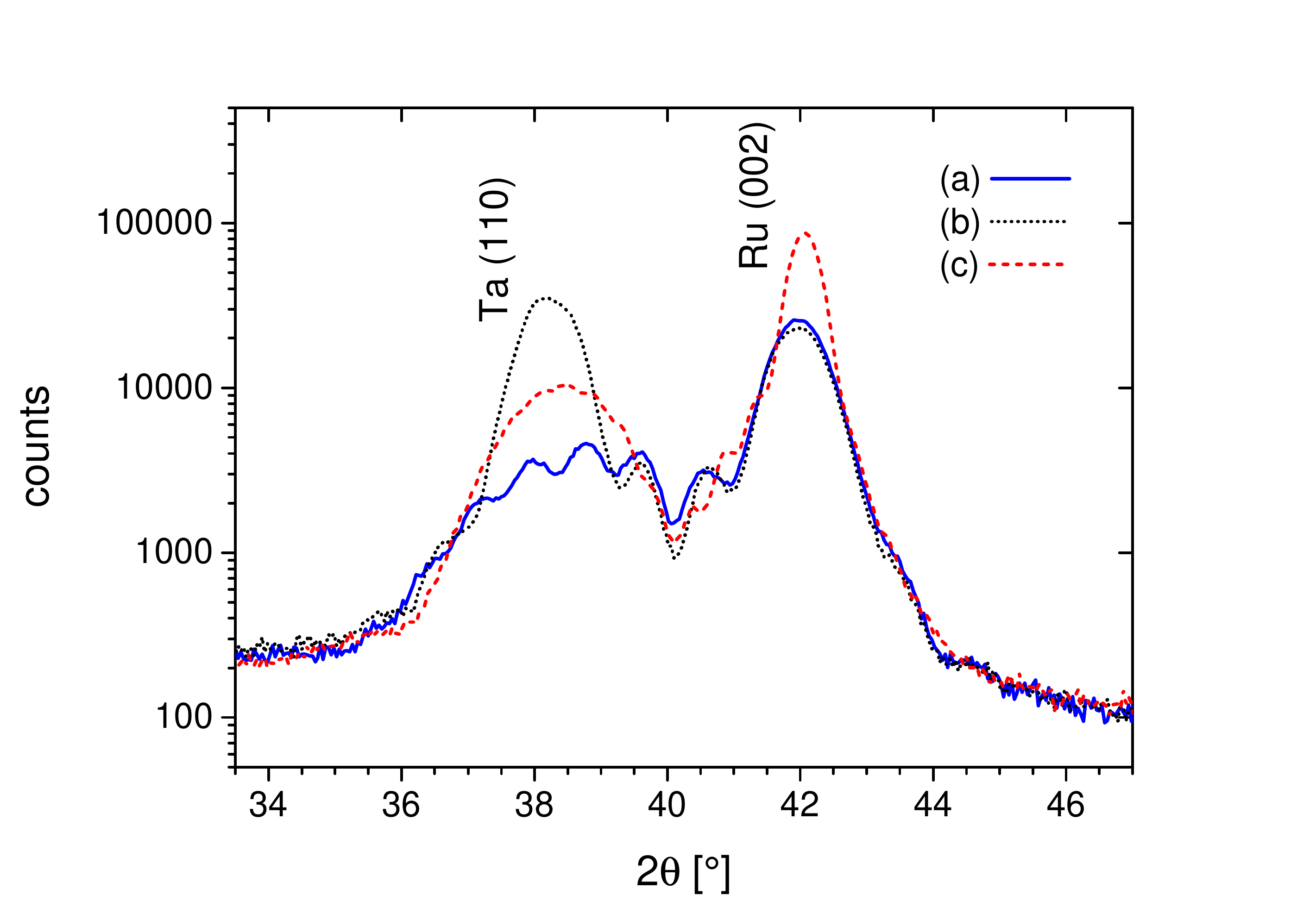}
            \caption{XRD $\theta$ - $2\theta$ profiles for each buffer before annealing.}
            \end{figure}
            
In order to determine the degree of the texture of the Ru and Ta layers in buffer systems (a), (b) and (c), the pole figures and rocking curves have been measured. Pole figures for buffer (b) at position $2\theta$ = 44$^{\circ}$, which corresponds to peak Ru (101) and at position $2\theta$ = 38.34$^{\circ}$, which corresponds to peak Ta (110) are shown in Fig.2.  
The upper pole figure shows a ring at position $\psi$ = 61.3$^{\circ}$, which is the angle between Ru (002) and Ru (101) planes. In the lower pole figure there is a spot in the center and a ring at position $\psi$ = 60$^{\circ}$, which is the angle between \{110\} planes in Ta.  Diffuse rings of Ru (101) and Ta (110) indicate that the layers have sheet texture with no crystallographic orientation in the layer plane. This confirms the fact that Ru and Ta buffer layers are polycrystalline with highly oriented columnar grains which contribute to roughness.     

            \begin{figure}[!htbp]
            \centering
            \includegraphics[width=0.5\textwidth]{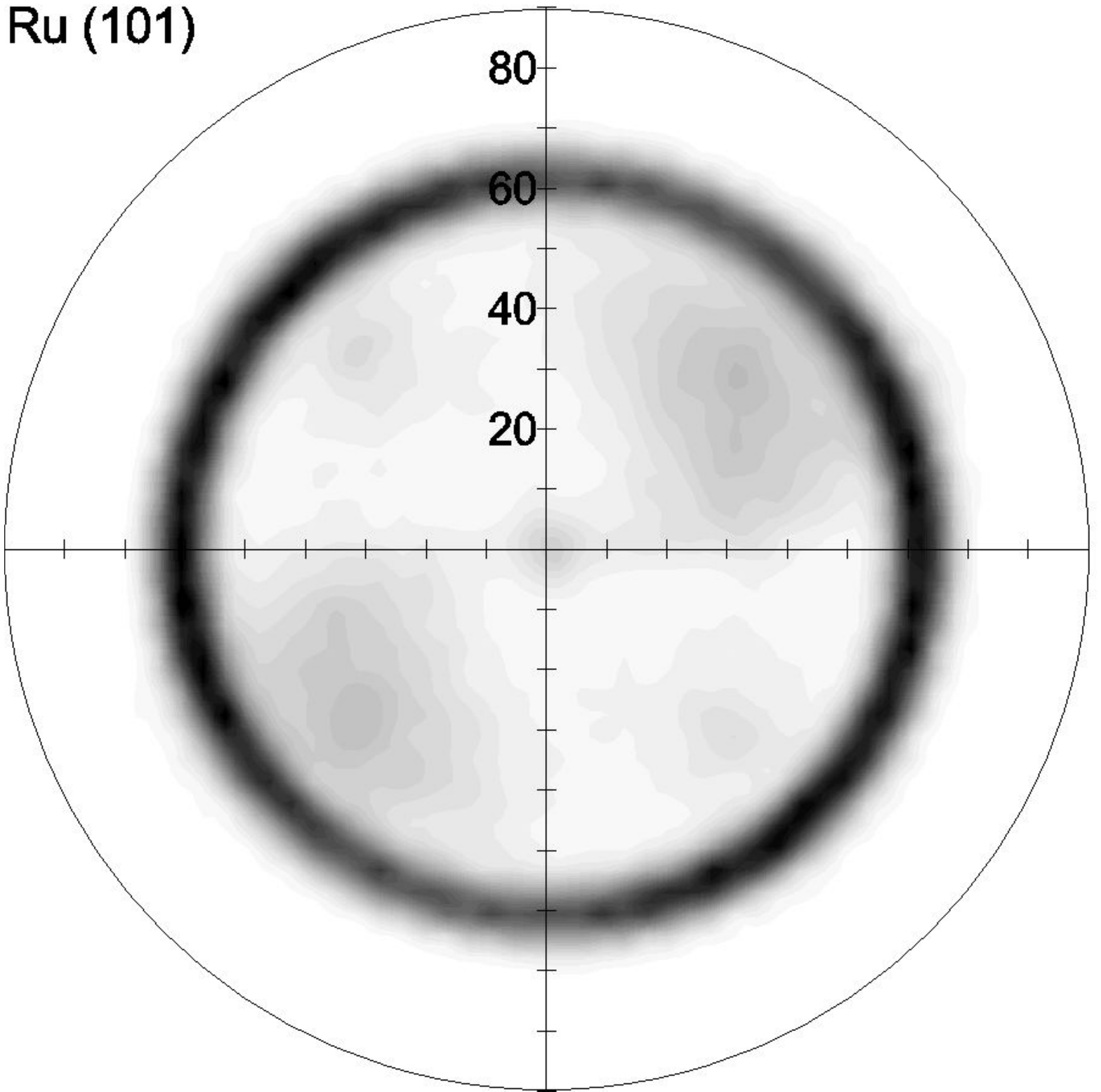}                            \includegraphics[width=0.5\textwidth]{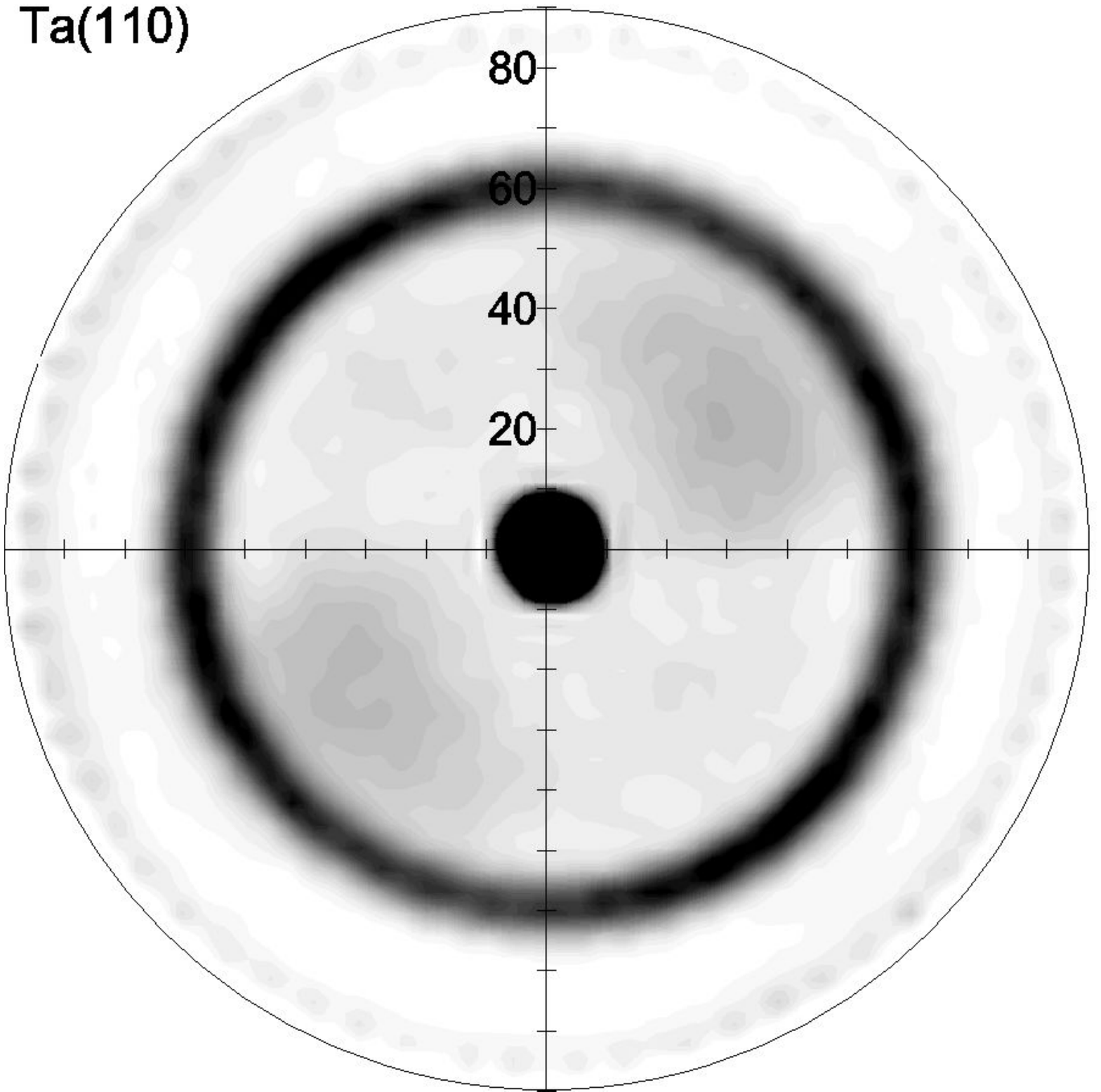}
            \caption{An example of XRD pole figures for Ru (101) and Ta (110) in the case of intermediately textured buffer (b).}
            \end{figure}

To verify the degree of texture for the buffers, we have measured rocking curve profiles on Ru (002) and Ta (110) peaks (Fig.3). In the case of Ru layers, the narrowest Full Width at Half Maximum (FWHM) was observed for buffer (c), which leads to conclusion that the Ru texture for this buffer is significantly higher than for buffers (a) and (b). The texture degree of the Ta layer deposited on Ru was shown to be the lowest for sample (a) and to increase in the case of samples (b) and (c). 

            \begin{figure}[!htbp]
            \centering
            \includegraphics[width=0.5\textwidth]{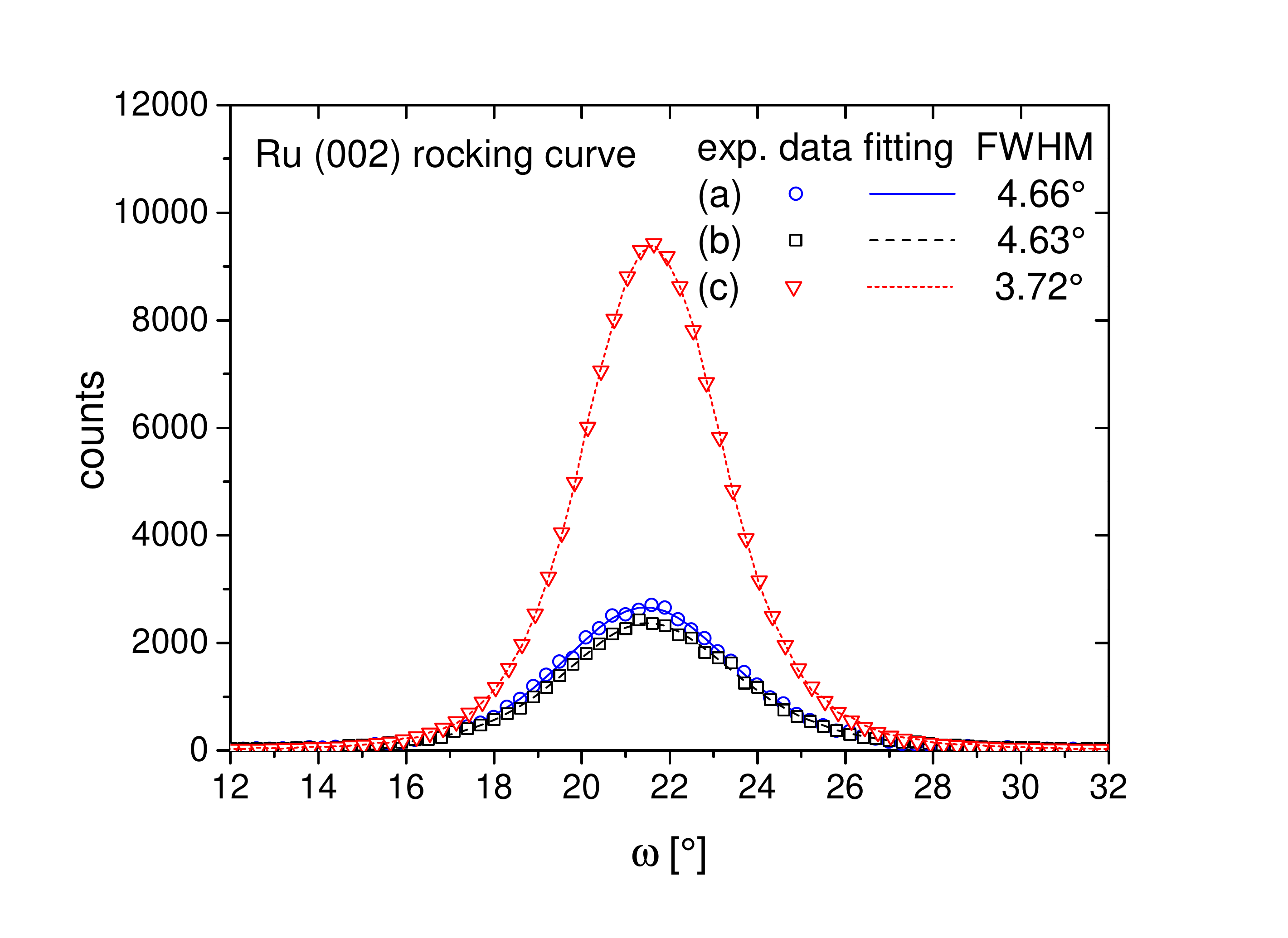}
           \includegraphics[width=0.5\textwidth]{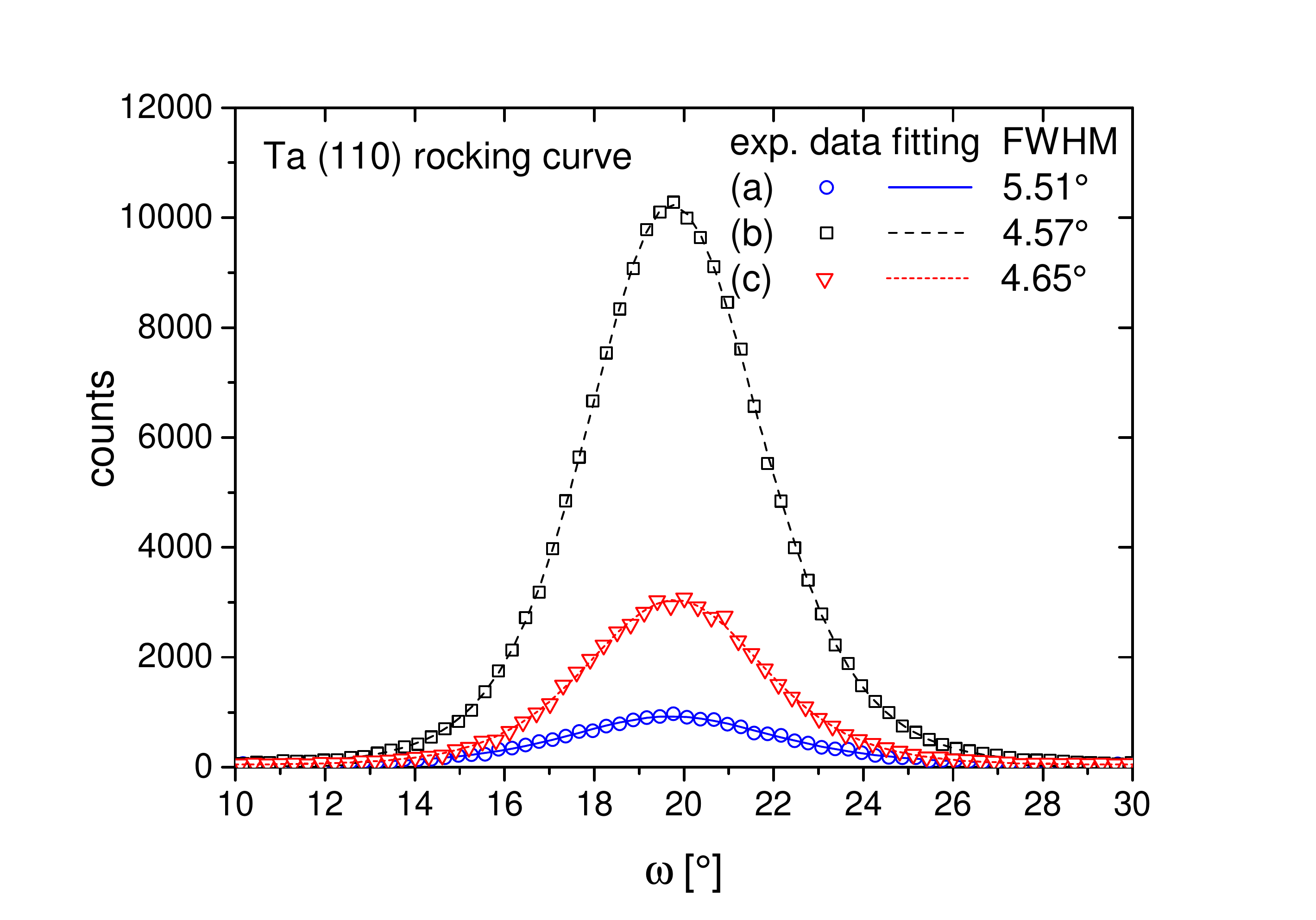}
            \caption{XRD rocking curve measurements for the Ru (002) and Ta (110) peaks for each buffer.}
            \end{figure}

Surface roughness of the single ferromagnetic layer systems with different buffers has been investigated with Atomic Force Microscopy (AFM) in a tapping mode. As-deposited single layer systems have been examined by scanning 500 nm x 500 nm areas on top of the structures. The Root Mean Square (RMS) roughness of stacks (a) and (b) with 10 nm of Ru in the buffer has been equal to 0.22 nm and 0.24 nm, respectively. %Ta layer covering Ru was fine crystalline ...... 
In the case of stack (c) with 20 nm of Ru, the roughness has increased to 0.28 nm. This is likely due to a highly oriented Ru polycrystalline columns on the lowermost amorphous Ta layer \cite{kanak2013x}. Resultant columnar grain diameters of the measured systems have been all at the same level of 15 nm, regardless of the type of the buffer. As the structural measurements have shown, buffer (a) has the smoothest surface while maintaining the weakest texture.

            \begin{figure}[!htbp]
            \centering
            \includegraphics[width=0.5\textwidth]{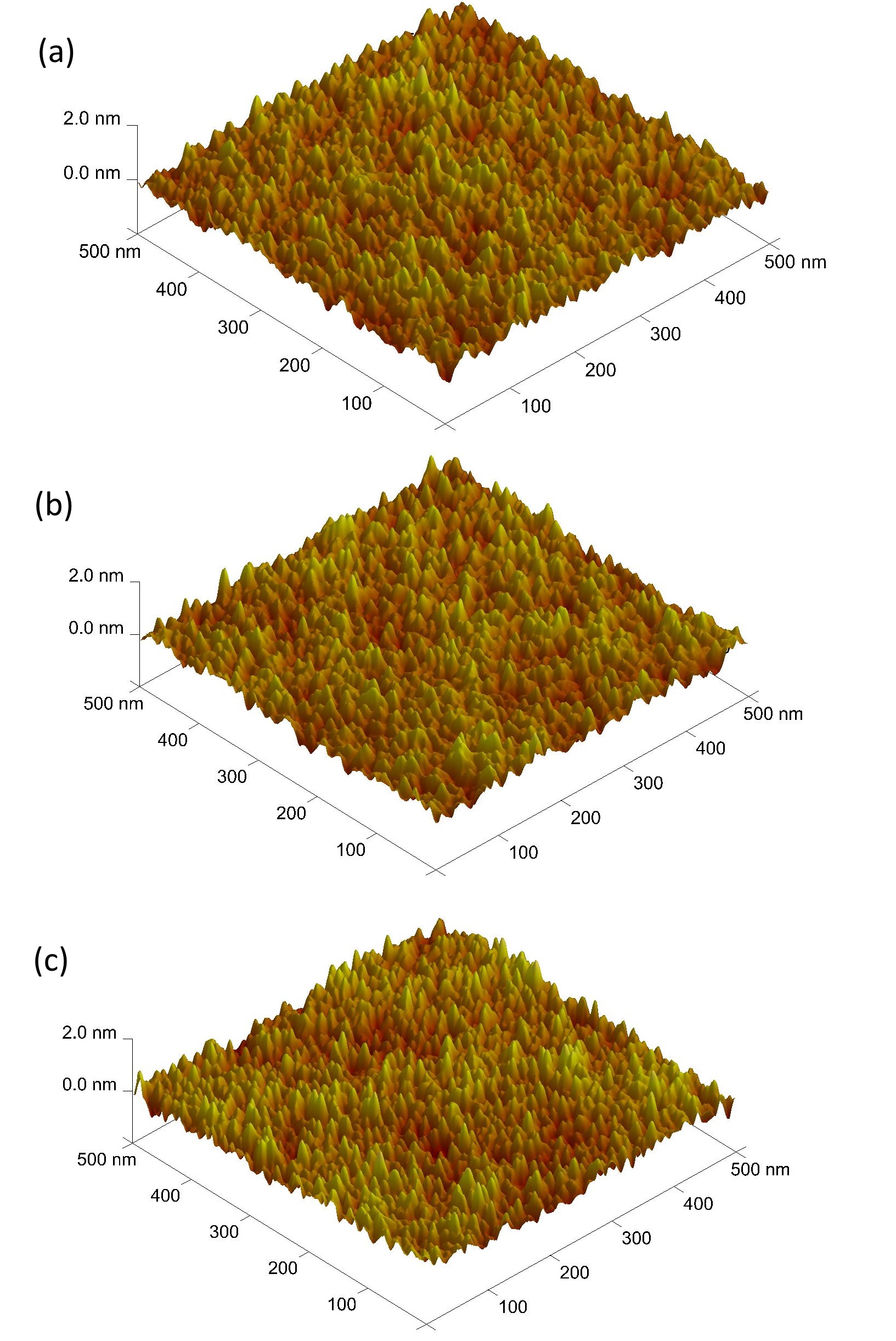}
            \caption{AFM images of the surface topography for single
ferromagnetic layer systems with buffers (a) RMS = 0.22 nm, (b) RMS = 0.24 nm, (c) RMS = 0.28, grains diameter is 15 nm.}
            \end{figure}

\subsection{Magnetic properties: dead layer and anisotropy}

The VSM measurement of the magnetic moment per unit area as a function of the nominal thickness for as-deposited and annealed structure with FeCoB \textit{bottom} layer is presented in Fig.5. 

            \begin{figure}[!htbp]
    	       \centering
    	       \includegraphics[width=0.5\textwidth]{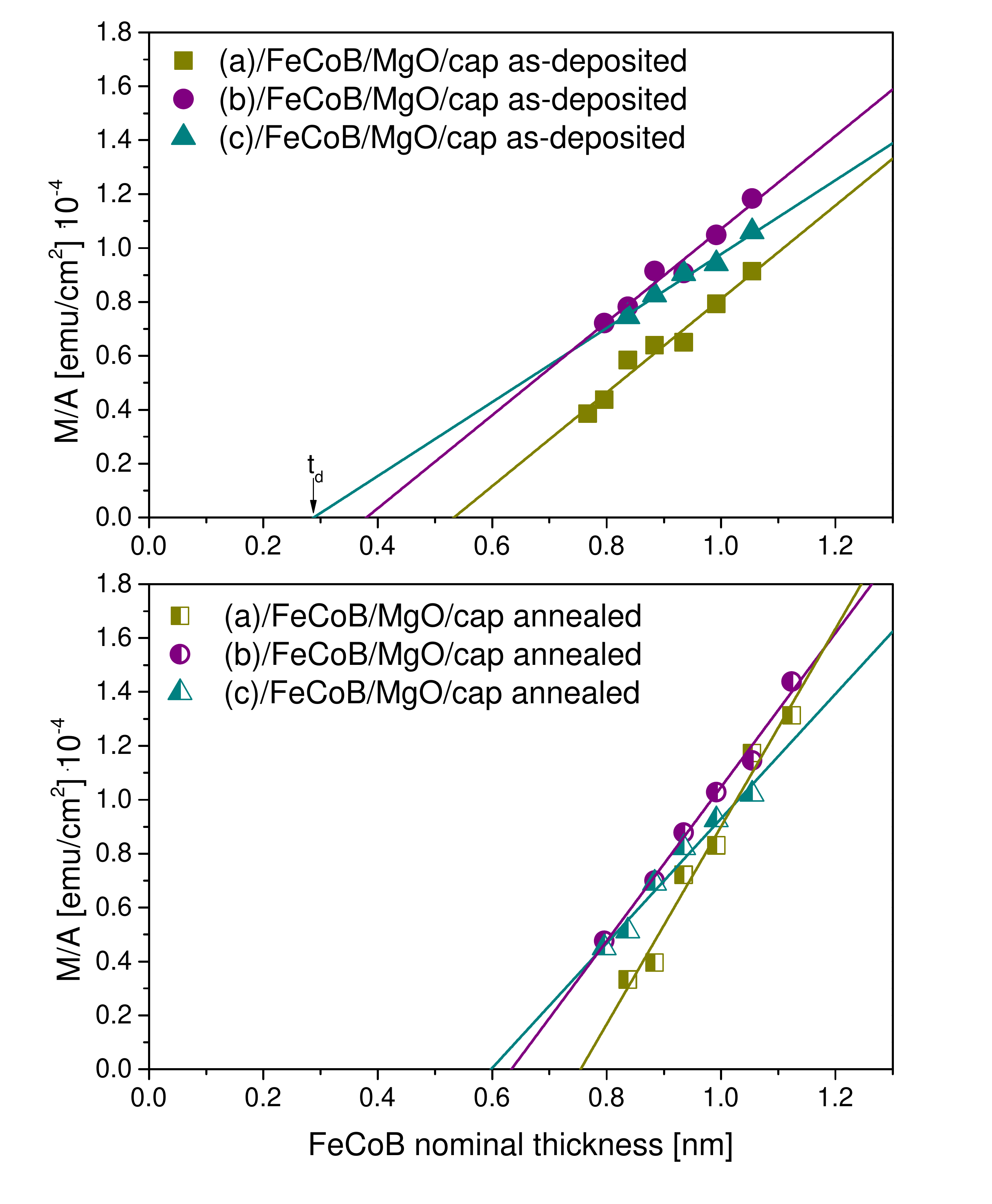}
    	       \caption{Magnetic moment per unit area of FeCoB layer in the as-deposited and annealed \textit{bottom} samples as a function of nominal thickness FeCoB with linear fits. The intersection point between the linear fit and the x axis is an estimated magnetic dead layer thickness t$_d$ for each buffer.}
            \end{figure}
            
            Using the intersection of the linear fit with the x axis, one can estimate the magnetic dead layer thickness t$_d$ for each buffer \cite{sinha2013enhanced}. %: (a): 0.54 nm, (b): 0.38 nm, (c): 0.29 nm. 
As seen in Fig.5, the largest dead layer thickness has been obseved for buffer (a) and the smallest one for buffer (c). %The magnetization for the annealed bottom structure is shown in Fig. 6. %The dead layer thickness for each buffer is now significantly bigger (a): 0.75 nm, (b): 0.64 nm, (c): 0.61 nm.

After annealing in magnetic field of 0.42 T, the dead layer thickness has increased in each sample. However, the character of their dependence on different buffer compositions has been preserved.             
Additionally, the domain structure for \textit{bottom} samples with FeCoB layer thickness of 1 nm has been examined by field-induced magnetization reversal process with a p-MOKE microscope (Fig.6). 

            \begin{figure}[!htbp]
    	       \centering
   	       \includegraphics[width=0.45\textwidth]{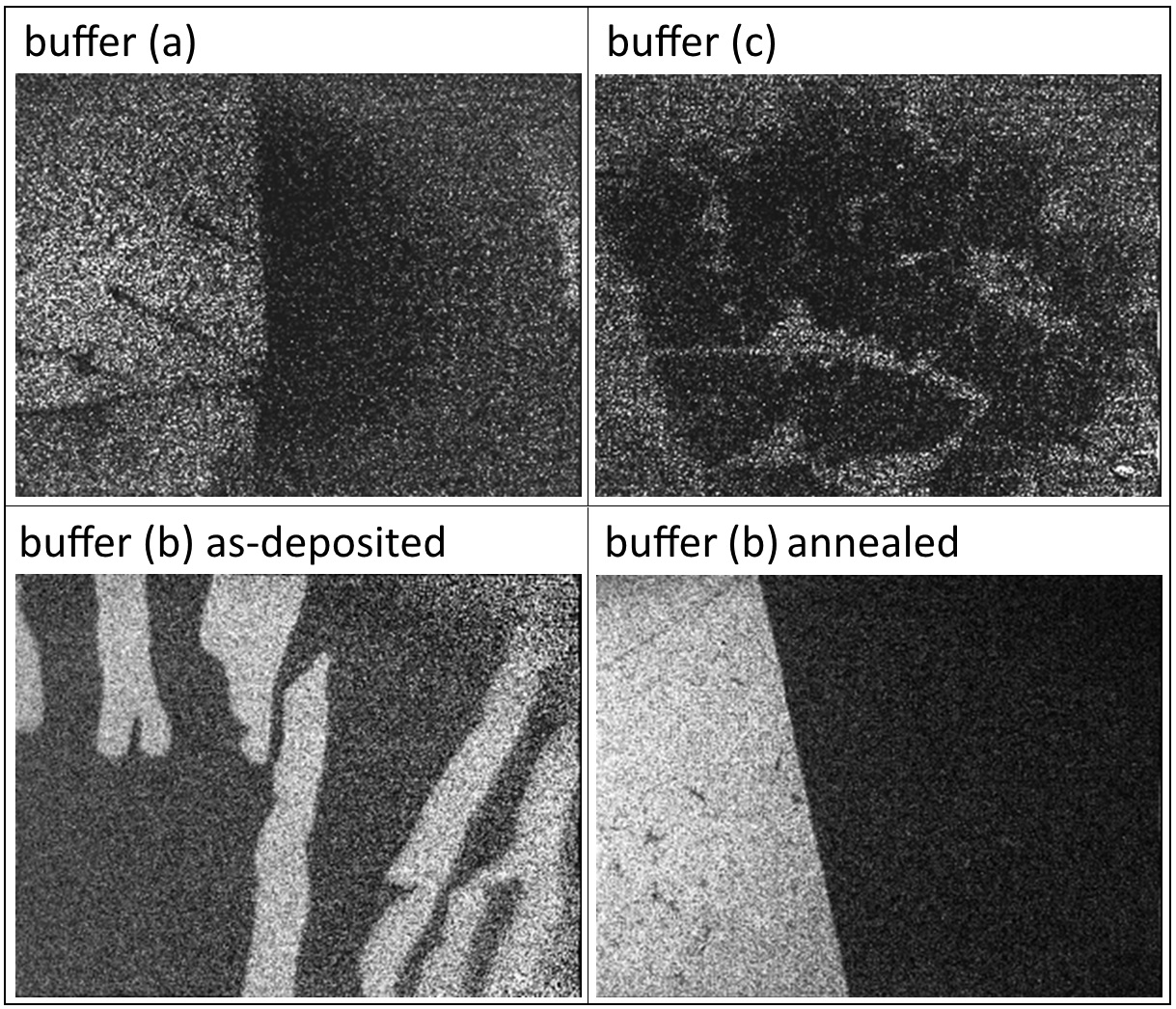}
    	       \caption{MOKE images during the magnetisation reversal of \textit{bottom} FeCoB layers for each buffer. Images for buffers (a) and (c) are taken from as-deposited samples.}
            \end{figure}

For both as-deposited and annealed structures with buffer (a), one large domain and a smooth domain wall propagation have been observed, in contrary to the irregular domains observed for samples with buffer (c). Such irregular domain structures are typical of films presenting a spatial disperson of the PMA energy barriers \cite{czapkiewicz2008thermally}. The intermediate results have been obtained in the case of buffer (b), for which the domain image acquired for the as-deposited sample shows stripe-like, irregular domains typical for samples with thickness near the spin reorientation transition regime. After annealing, the domain size increased, indicating a uniform spin orientation perpendicular to the plane. This effect may be caused directly as a result of the atomic ordering during annealing in external magnetic field or indirectly, by changing the effective thickness of the ferromagnetic layer through the enlargement of the paramagnetic dead-layer. Because the annealing has kept a monotonic change of the FeCoB parameters for buffers (a), (b), (c), below we focus only on measurements taken for the extreme cases (a) and (c). Figure 7 shows the magnetic moment per unit area of \textit{top} samples before and after the annealing. Clearly, the magnetic dead layer thickness in the case of \textit{top} samples is higher than in the \textit{bottom} ones. This tendency can be explained by greater interface diffusion when Ta capping layer is sputtered on FeCoB for \textit{top} structures, in contrast to \textit{bottom} structures, where FeCoB is deposited on the Ta sublayer of the buffer. We assume that the mixing at the Ta-FeCoB interfaces is in general induced by a large negative interfacial enthalpy. Therefore, the larger absolute value of mixing enthalpy for Ta in Fe (-67 kJ/mole of atoms) and for Ta in Co (-109 kJ/mole of atoms) compared to that for Fe in Ta (-54 kJ/mole of atoms) and Co in Ta (-86 kJ/mole of atoms) \cite{boer1988cohesion} may partially explain the difference in dead layers thicknesses. Even more important may be the fact  that during the magnetron sputtering of Ta on FeCoB, heavy Ta atoms penetrate the FeCoB layer more easily in one case than the other.

            \begin{figure}[!htbp]
    	       \centering
   	       \includegraphics[width=0.5\textwidth]{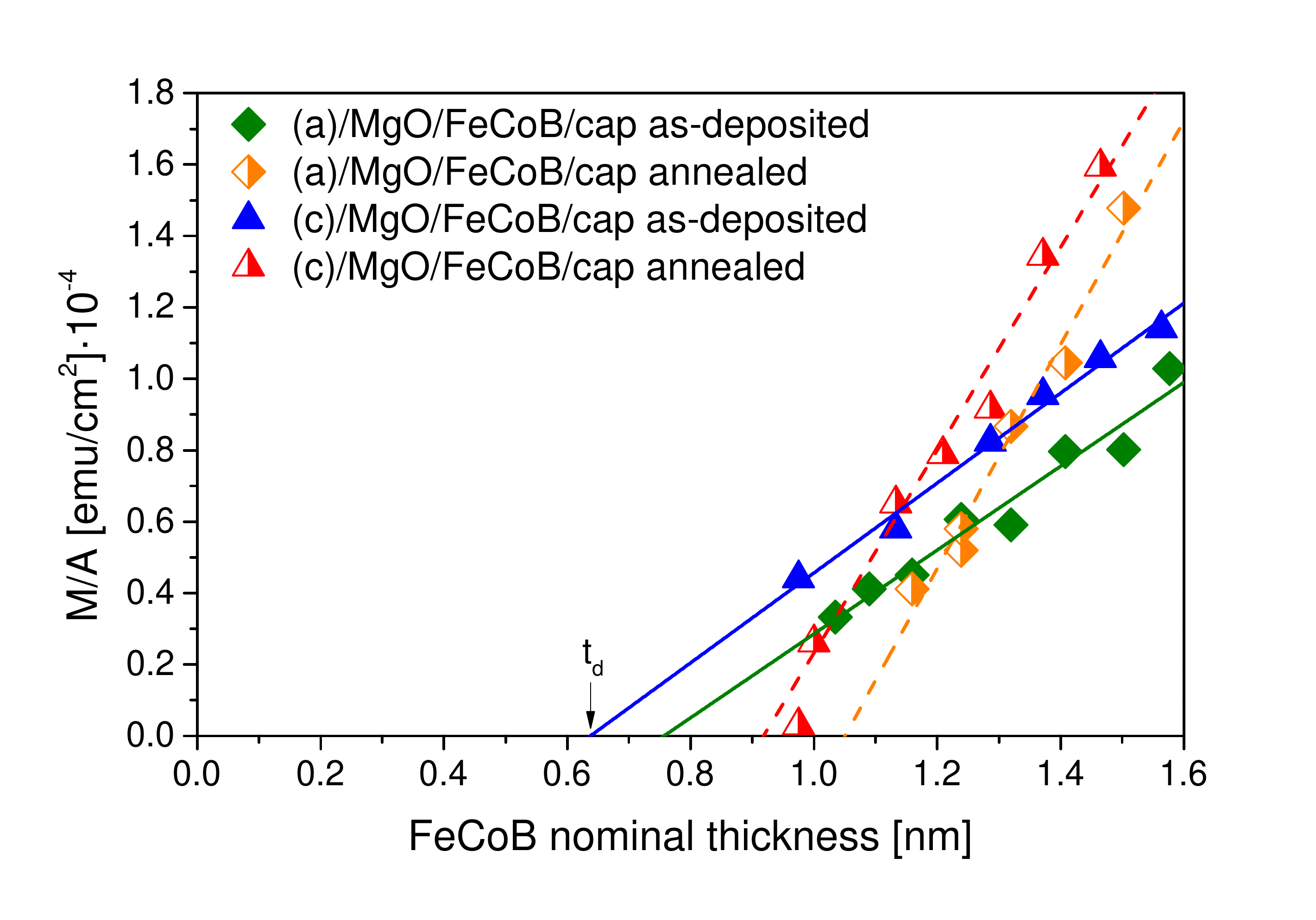}
    	       \caption{Magnetic moment per unit area of FeCoB layer in the annealed \textit{top} sample as a function of nominal thickness FeCoB with linear fits.}
            \end{figure}
            
VSM measurements have shown a strong PMA in single FeCoB layer \textit{bottom} systems and weak in-plane anisotropy in the \textit{top} layer case. For double layer systems with MgO thickness of 1 nm, both anisotropies have preferred magnetization vectors orientation perpendicular to the sample plane. The anisotropy field $H_k$ for buffer (a) has been equal to 1010 Oe (FeCoB above MgO) and 5620 Oe (FeCoB below MgO), while for buffer (c) it has been equal to 920 Oe (FeCoB above MgO) and 5330 Oe (FeCoB below MgO). One can see that the $H_k$ value is significantly higher in the case of FeCoB below MgO system, similarly to the single layer measurements. Additionally, buffer (a) has higher $H_k$ values than buffer (c).

\subsection{Critical current and thermal stability}

In order to perform CIMS experiment we have nano-patterned annealed FeCoB bilayers with MgO thickness of 1 nm and resistance-area product equal to 40 $\Omega \mu$m$^2$. We note that, although for the chosen barrier thickness FeCoB layer above MgO has appeared to be near the spin reorientation transition region, for bilayer nano-pillars with small planar dimensions (between 100 and 200 nm) the shape anisotropy gives smaller contribution to the in-plane anisotropy component and therefore strong effective PMA is observed (inset in Fig.8). 

            \begin{figure}[th]
    	       \centering
    	       \includegraphics[width=0.5\textwidth]{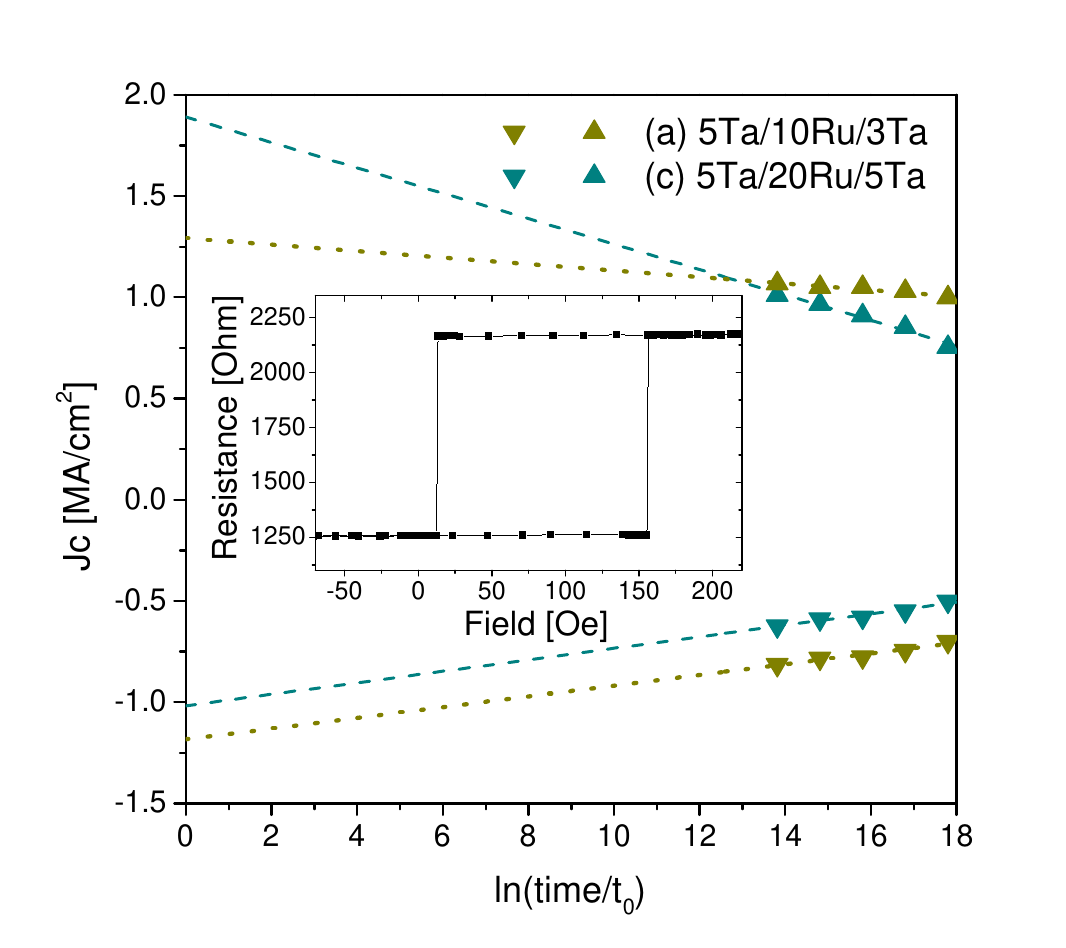}
    	       \caption{Critical current densities for buffers (a) and (c) in function of different pulse time widths. t$_0$ was equal to 1 ns. The perpendicular magnetic field hysteresis loop for the sample on the (a) buffer (inset).}
            \end{figure} 

An example of field hysteresis loop is depicted in the inset of Fig.8. We have used magnetic field bias to compensate the hysteresis field shift in order to perform a CIMS experiment. From the current polarity, we can identify the layer above the MgO barrier to be a free layer.
The CIMS hysteresis loop has been measured for different time pulse widths in order to estimate the intrinsic critical current \cite{kubota2005evaluation} (Fig.8). We have obtained slightly better critical current values for the buffer (a) with the thickest dead layer: $J_{crit}^+=1.3$ MA/cm$^2$, $J_{crit}^-=1.2$ MA/cm$^2$ and $J_{crit}^{avg}=1.25$ MA/cm$^2$ than for the buffer (c) with a thin dead layer: $J_{crit}^+=1.9$ MA/cm$^2$, $J_{crit}^-=1.1$ MA/cm$^2$ and $J_{crit}^{avg}=1.5$ MA/cm$^2$. However, the calculated thermal stability factors for the junctions are $\Delta^{+}=35$, $\Delta^{-}=30$ and $\Delta^{avg}=32.5$ for buffer (c), while for buffer (a) we obtained $\Delta^{+}=81$, $\Delta^{-}=45$ and $\Delta^{avg}=63$. We note that the last value is greater than the commonly assumed limit of 40 \cite{ikeda2007magnetic}. In other words, the sample (a) not only preserves a desirably low critical current, but even further decreases it while greatly enhancing the thermal stability. 
%In other words, the sample (a) greatly enhances the thermal stability and not only preserves a desirably low critical current, but even further decreases it. 

However, the fact that sample (a) has greater thermal stability factor than sample (c) while maintaining similarly small $J_{crit}$ demands an explanation.
We believe that such behaviour can be accounted for the decrease of the damping coefficient in the sample (a) that compensates for the increase of the energy barrier needed for the STT switching. In order to confirm this hypothesis, we have conducted VNA-FMR measurements on the annealed single layer systems and calculated the damping factor $\alpha$, using the standard formula \cite{rossing1963resonance}:

\begin{equation}
\Delta H=\Delta H_0+\alpha\frac{4\pi f}{\gamma},
\end{equation}
where $\Delta H_0$ is a frequency-independent component of
line width which originates from magnetic inhomogeneities and $\gamma$ is the gyromagnetic ratio. Results are presented in Fig.9. 

            \begin{figure}[th]
    	       \centering
    	       \includegraphics[width=0.5\textwidth]{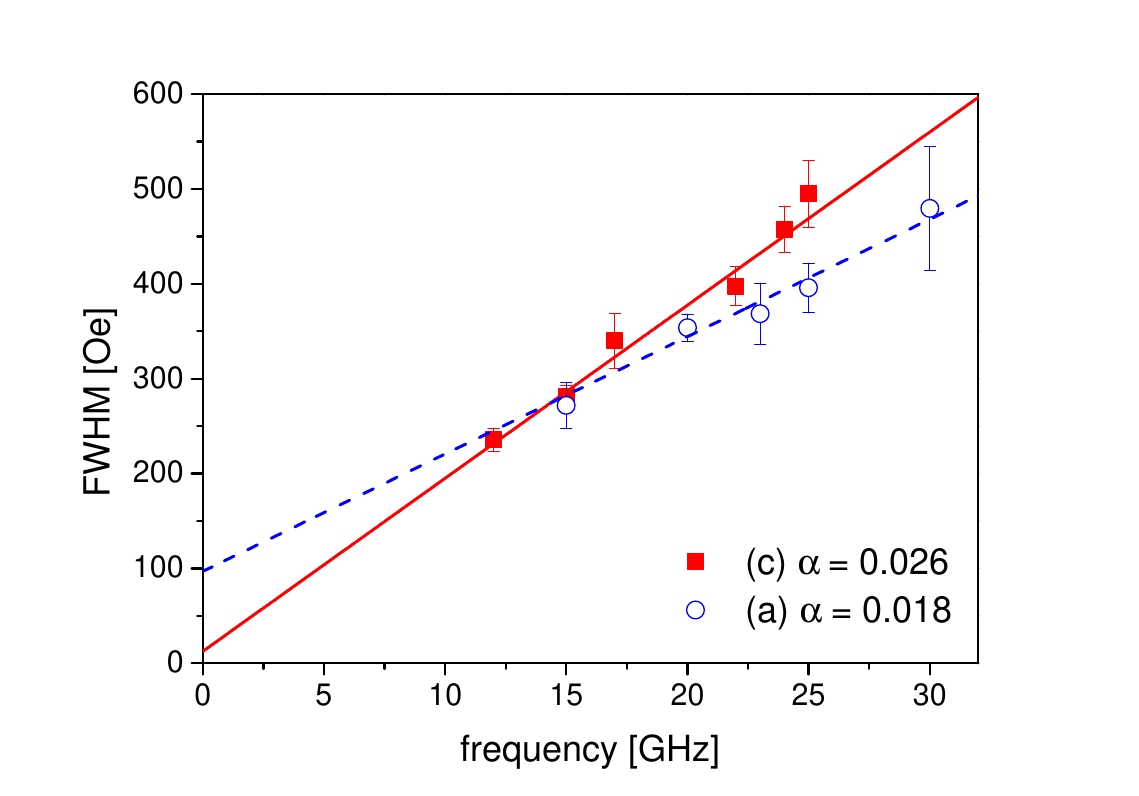}
    	       \caption{VNA-FMR measurement for the annealed \textit{top} layers on buffers (a) and (c).}
            \end{figure}

One can see that the measurement is in agreement with our hypothesis, as the damping factor for buffer (c) $\alpha=0.026$ is 44\% greater than the factor for buffer (a) $\alpha=0.018$. Regarding the results of the structural measurements, we suppose that the smaller damping originates from the smoother interface of the layers deposited on buffer (a) \cite{serrano2011inductive}.

\section{Summary and conclusions\label{sec:summary}}
We have investigated the influence of three different buffers on the properties of MTJs with PMA. The thinnest magnetically dead layer has been observed for buffer Ta 5 / Ru 20 / Ta 5, which has the strongest texture, the biggest roughness and produces irregular domain images. The thickest dead layer has been observed for buffer Ta 5 / Ru 10 / Ta 3, which has the weakest texture, the smallest roughness and produces MOKE images with one large domain. Buffer Ta 5 / Ru 10 / Ta 10 exhibited intermediate properties between the other two. What is more, we have shown that the anisotropy fields for buffer Ta 5 / Ru 10 / Ta 3 are larger than for buffer Ta 5 / Ru 20 / Ta 5.

By means of CIMS experiments we have obtained the critical current values of 1.25 MA/cm$^2$ for Ta 5 / Ru 10 / Ta 3 and of 1.5 MA/cm$^2$ for Ta 5 / Ru 20 / Ta 5. However, there is a two-fold difference in thermal stability factors between these two buffer structures. Buffer Ta 5 / Ru 10 / Ta 3 produces $\Delta$ equal to 63. We have shown that the rough buffer with a strong texture has damping factor 44\% greater than the smooth one. We conclude that the difference in damping factors compensates for the difference in the switching barrier heights. As a result, by adjusting buffer characteristics one can obtain a significant increase in thermal stability factors while keeping the critical current values at a similar level. This can be important for the further optimization of the MTJs. 

%Therefore, the buffer layers which induce the interfacial roughness through the texture affect the damping factor value and should be chosen carefully in order to optimize the critical current and thermal stability in MTJs.

\section*{Acknowledgements}
M.F. and J.Ch. acknowledge Polish Ministry of Science and Higher Education Diamond Grant DI 2011001541. A.Ż., M.B., J.D, H.G, J-Ph.A. and T.S. acknowledge the NANOSPIN Grant no. PSPB-045/2010 from Switzerland through the Swiss Contribution. J.K., M.B., M.C. and W.P. acknowledge Polish National Science Center Grant DEC-2012/05/E/ST7/00240. W.S. acknowledges support from Foundation for Polish Science through START programme.
\bibliographystyle{nature}

\bibliography{bibliography}

\end{document}